\documentclass[proof]{WileyASNA-v1}

\articletype{}%

\received{}
\revised{}
\accepted{}

\begin{document}

\title{Exploring Dark Photon Production and Kinetic Mixing Constraints in Heavy-Ion Collisions}

\author[1,2,3]{Adrian William Romero Jorge*}
\author[4,3,2]{Elena Bratkovskaya}
\author[4]{Taesoo Song}
\author[2]{Laura Sagunski}
\authormark{Adrian William Romero Jorge}
%
%
\address[1]{ Frankfurt Institute for Advanced Studies, Ruth Moufang Str. 1, 60438 Frankfurt, Germany}
\address[2]{Institute for Theoretical Physics, Johann Wolfgang Goethe University, Max-von-Laue-Str. 1, 60438 Frankfurt am Main, Germany}
\address[3]{Helmholtz Research Academy Hessen for FAIR (HFHF), GSI Helmholtz	Center for Heavy Ion Physics. Campus Frankfurt, 60438 Frankfurt, Germany}
\address[4]{GSI Helmholtzzentrum für Schwerionenforschung GmbH, Planckstr. 1, 64291 Darmstadt, Germany}
\corres{GSI Helmholtzzentrum fuer Schwerionenforschung GmbH, Planckstrasse 1, 64291 Darmstad, Germany. KBW 2.08 -\email{E.Bratkovskaya@gsi.de}}


\abstract{
Vector $U$-bosons, often referred to as 'dark photons', are potential candidates for mediating dark matter interactions. In this study, we outline a procedure to derive theoretical constraints on the upper bound of the kinetic mixing parameter $\epsilon^2(M_U)$ using dilepton data from heavy-ion  from SIS to RHIC energies. The analysis is based on the microscopic Parton-Hadron-String Dynamics (PHSD) transport model, which successfully reproduces the measured dilepton spectra in $p+p$, $p+A$, and $A+A$ collisions. Besides the dilepton channels resulting from interactions and decays of Standard Model particles (such as mesons and baryons), we extend the PHSD approach to include the decay of hypothetical $U$-bosons into dileptons, $U \to e^+ e^-$. The production of these $U$-bosons occurs via Dalitz decays of pions, $\eta$-mesons, $\omega$-mesons, Delta resonances, as well as from the decays of vector mesons and $K^+$ mesons. This analysis provides an upper limit on $\epsilon^2(M_U)$ and offers insights into the accuracy required for future experimental searches for dark photons through dilepton experiments.}


\maketitle

\section{Introduction}
The existence of dark matter (DM) is strongly supported by astrophysical and cosmological observations, one of the most compelling being the rotation curves of galaxies. These curves, which plot the total orbital velocities 
as a function of their distance from the galactic center, remain flat at large radii, defying expectations based on the visible distribution of matter. This discrepancy suggests the presence of an unseen mass providing direct evidence for a dark matter halo enveloping galaxies ~\citep{Bertone:2016nfn,Tulin:2017ara}. Gravitational lensing also provides evidence, where light from distant objects is bent by massive foreground structures, indicating the presence of dark matter. These phenomena emphasize the importance of dark matter in shaping the Universe's structure, from the smaller scales to the large-scale structure although its fundamental properties remain unknown.

It has been proposed that mediators of dark matter can interact with Standard Model (SM) particles through four potential "portals" — vector, Higgs, neutrino, and axion portals (see \cite{Alexander:2016aln,Battaglieri:2017aum,Agrawal:2021dbo} for reviews and references).

The vector portal suggests the existence of a gauge symmetry group mixing between $U(1)$ and $U(1)^\prime$ \cite{Holdom:1985ag}, where the Lagrangian includes the hypercharge field-strength tensor of the SM photon field and that of the dark matter  vector boson field: ${\cal L} \sim \epsilon^2/2 , F_{\mu\nu}{F^{\mu\nu}}^\prime$. Mediators in this framework are vector bosons, often referred to as $U$-bosons, "dark photons," "hidden photons," or $A^\prime$. The mass of these $U$-bosons, $M_U$, remains unknown. The kinetic mixing parameter $\epsilon^2$ quantifies the interaction strength between SM and DM particles \cite{Fayet:1980ad,Fayet:2004bw,Boehm:2003hm,Pospelov:2007mp,Batell:2009di,Batell:2009yf}. This mixing allows $U$-bosons to decay into lepton pairs such as $e^+e^-$ or $\mu^+\mu^-$.

$U$-bosons can be produced via the Dalitz decay of SM particles, such as pseudoscalar mesons ($\pi^0$ and $\eta$), baryonic resonances like $\Delta$'s, and other possible decay channels. These processes enable the potential detection of dark photons in dilepton experiments, a prospect that has spurred significant experimental and theoretical interest \cite{Alexander:2016aln,Battaglieri:2017aum,Beacham:2019nyx,Billard:2021uyg}. For instance, the excess electronic recoil events observed by the XENON1T Collaboration could be interpreted as evidence of dark matter sources including dark photons as plausible candidates \citep{Aprile:2020tmw}.

 Key experimental constraints on dark photons come from beam dump experiments, fixed-target setups, collider searches, and rare meson decays, complemented by cosmological and astrophysical bounds related to the cosmic microwave background and stellar cooling ~\citep{Fabbrichesi:2020wbt}.
Searches for dark photons have systematically tightened the exclusion limits on the kinetic mixing parameter $\epsilon^2$, achieving sensitivities around $10^{-6}$ for masses between 20 MeV and a few GeV. Key contributions have come from fixed-target experiments like A1 \cite{Merkel:2014avp}, NA48/2 \cite{Batley:2015lha}, and APEX \cite{APEX:2011dww}, as well as collider-based searches such as BaBar \cite{BaBar:2009lbr, BaBar:2014zli} and LHCb, which explored higher mass ranges with remarkable precision. These results, complemented by limits from experiments like KLOE \cite{KLOE-2:2014qxg} and CMS \cite{CMS:2023hwl}, collectively map the parameter space, providing stringent constraints on the existence of dark photons.

This study builds upon previous works (cf. ~\citep{Schmidt:2021hhs,Bratkovskaya:2022cch}) that investigated the upper limit of the kinetic mixing parameter $\epsilon^2(M_U)$ for hypothetical $U$-bosons with a mass $M_U \leq 0.6$ GeV, extending the analysis to cover a mass range up to 2 GeV. 
Beyond the previously examined Dalitz decays of $\pi^0$, $\eta$, and $\Delta$ particles, this study introduces additional dark photon production channels, including the Dalitz decay of the $\omega$, direct decays of vector mesons such as the $\rho$, $\omega$, and $\phi$, as well as decays from $K^+$ mesons.

\section{Standard matter production in the PHSD  Approach}

The Parton-Hadron-String Dynamics (PHSD) approach is a microscopic transport model that operates in a non-equilibrium regime, incorporating both hadronic and partonic degrees of freedom. It describes the full evolution of relativistic heavy-ion collisions from the initial hard nucleon-nucleon (NN) collisions to the formation of the quark-gluon plasma (QGP), its subsequent partonic interactions, and the final hadronization and hadronic interactions of the produced particles. The dynamical evolution of the system is governed by the Cassing-Juchem off-shell transport equations in the test-particle representation, derived from the Kadanoff-Baym equations, which provide a robust description of strongly interacting systems in nonequilibrium states. These transport equations are used to model both the partonic and hadronic phases of the collision dynamics, allowing for a seamless transition between them \cite{Cassing:2008sv,Cassing:2008nn,Cassing:2009vt,Bratkovskaya:2011wp,Konchakovski:2011qa}.

The hadronic sector of PHSD builds upon the Hadron-String Dynamics (HSD) approach, incorporating a wide range of hadronic states such as baryons, mesons, and higher resonances. Multi-particle production is modeled via the Lund string model, adjusted for intermediate-energy reactions and medium effects. Strings, formed in primary and secondary interactions, decay into hadrons or pre-hadrons, which follow formation time dynamics.

Hadronization occurs at an energy density of $\epsilon_C \sim 0.5$ GeV/fm³, consistent with lattice QCD. In the QGP phase, partons evolve according to the DQPM, using temperature- and density-dependent quasi-particles. The QGP transitions to hadrons as it nears the hadronization temperature, with final hadronic interactions modeled by off-shell HSD dynamics, including optional self-energies.

The PHSD approach has been successfully applied to collisions from SIS18 to RHIC energies, accurately reproducing many experimental observables. This success underscores its versatility and reliability in describing a wide range of phenomena in high-energy nuclear physics \cite{Cassing:2008sv,Cassing:2008nn,Cassing:2009vt,Bratkovskaya:2011wp,Konchakovski:2011qa,Linnyk:2015rco,Moreau:2019vhw,Song:2018xca}.

\subsection{Dilepton production from the SM sources in the PHSD}

Dileptons ($e^+e^-$, $\mu^+\mu^-$ pairs) produced from virtual photon decay can be emitted from all stages of heavy-ion reactions, originating from both hadronic and partonic sources \cite{Rapp:2013nxa, Linnyk:2015rco}. At low invariant masses ($M < 1$ GeV), hadronic sources include Dalitz decays of mesons and baryons (e.g., $\pi^0$, $\eta$, $\Delta$) and direct decays of vector mesons ($\rho$, $\omega$, $\phi$), as well as hadronic bremsstrahlung. For intermediate masses (1 GeV $< M < 3$ GeV), dileptons arise from semileptonic decays of $D+\bar{D}$ pairs and multi-meson reactions (e.g., $\pi+\pi$, $\pi+\rho$, $\pi+\omega$). At high masses ($M > 3$ GeV), sources include the direct decay of heavy vector mesons like $J/\Psi$, $\Psi^\prime$, and hard Drell-Yan annihilation. Partonic sources involve QGP dileptons, primarily from thermal $q\bar{q}$ annihilation and Compton scattering in the QGP phase, which are most significant in the intermediate mass regime.

This study focuses on low-energy heavy-ion collisions, thus only hadronic sources are relevant. For QGP-related dilepton production the reader is referred to \cite{Linnyk:2015rco} and \cite{Song:2018xca}.

Dilepton production from baryonic or mesonic resonance decays is described by the following processes:
\begin{align}
    & BB \to R X, \quad \quad mB \to R X, \quad \quad 
      mm \to R X,\\ 
     &  R \to e^+e^-, \quad  \quad  R \to e^+e^- X, \\
    & R \to m X, \ m \to e^+e^- X,  \quad 
     R \to R' X, \ R' \to e^+e^- X.
\end{align}
Resonances $R$ can be produced in baryon-baryon, meson-baryon, or meson-meson collisions, or at high energies during the hadronization process. These resonances can decay directly to dileptons or via Dalitz decay. Some resonances may decay to other resonances, which then decay into dileptons. The electromagnetic contributions of conventional dilepton sources (e.g., $\pi^0$, $\eta$, $\omega$, $\Delta$ Dalitz decays and direct decays of vector mesons) are calculated based on previous work \cite{Bratkovskaya:2000mb}.
%
%
%
\section{Dark photon production and dilepton decay channels  in the PHSD}
In this study we adopt the methodology developed in prior research ~\citep{Schmidt:2021hhs}. Alongside the previously considered Dalitz decays of $\pi^0, \eta $ and $\Delta$-resonances,
\begin{eqnarray}
\pi^{\mathbf{0}}, \eta \rightarrow \gamma U, \quad \quad
\Delta \rightarrow N U. 
\end{eqnarray}
Additionally, we consider further production channels for dark photons: direct decays of vector mesons, specifically $V = \rho, \phi, \omega$ as well as the  Dalitz decay of $\omega$ and the kaon decay,
\begin{eqnarray}
\rho, \boldsymbol{\phi}, \omega \rightarrow U, \quad \quad
\boldsymbol{\omega} \rightarrow \boldsymbol{\pi}^{\mathbf{0}} U,  \quad \quad
K^{+} \rightarrow \boldsymbol{\pi}^{+} U.
\end{eqnarray}
The dilepton yield from a $U$-boson decay of mass $M_U$ can be evaluated as the sum of all possible contributions 
for a given mass $M_U$:
\begin{eqnarray}
N^{U\to e^+e^-} = 
\sum_{h} N^{U\to e^+e^-}_{h}  
 = Br^{U\to e^+e^-} 
 \sum_{h}
 N_{h \to  U X},\label{NUee}
\end{eqnarray}
\noindent
where $Br^{U\to e^+e^-}$ is the branching ratio for the decay of $U$-bosons to $e^+e^-$ and, $X = \gamma$ for $h = \pi, \eta, \omega$, $X = N$ for $h = \Delta$, and $X = \pi$ for $h = \omega, K^+$. 
We assume that the width of the $U$-boson is zero (or very small), i.e., it contributes only 
to a single $dM$ bin of dilepton spectra from SM sources. 
On the other hand,  the yield of $U$-bosons  of mass $M_U$ themselves can be
estimated from the coupling to the virtual photons ~\citep{HADES:2013nab}:
\begin{eqnarray}
\footnotesize{
 N_{h\to UX} = N_h Br_{h\to \gamma X}  \cdot    
   \frac{\Gamma_{h\to UX}}{\Gamma_{h\to \gamma X}}, }\label{mNU2}
\end{eqnarray}
where $X=\pi,\eta$ for $h=\gamma$ and $X=\Delta$ for $h=N$. The ratio of the partial widths for the Dalitz decays of $\pi^0$ and $\eta$ mesons into $U$-bosons and real photons can be evaluated, as taken from Refs. ~\citep{Batell:2009yf,Batell:2009di}.
For the evaluation of the partial decay widths of a broad $\Delta$  resonance, one has to take into account the $\Delta$ spectral function $A(M_\Delta)$ as used also in the HADES study ~\citep{HADES:2013nab}. 
For other dark photon decays with $h=V$, $\omega$, and $K^+$, the partial widths are taken from the models developed in Refs. ~\citep{Batell:2009di}, ~\citep{Gorbunov:2024nph}, and ~\citep{Pospelov:2008zw}, respectively,
\begin{align}
&    Br(P \rightarrow \gamma U)=\varepsilon^2 B r(P \rightarrow \gamma \gamma)\left(1-\frac{M_U^2}{m_P^2}\right)^3,  
    {\tiny P=\pi, \eta, }\\
& B r(\Delta \rightarrow N U)=\varepsilon^2 B r(\Delta \rightarrow N \gamma) \int A\left(m_{\Delta}\right) \times \nonumber \\
& \hspace{5cm} \frac{\lambda^{3 / 2}\left(m_{\Delta}, m_N, M_U\right)}{\lambda^{3 / 2}\left(m_{\Delta}, m_N, 0\right)}, 
\end{align}

\begin{align}
& B r\left(\omega \rightarrow \pi^0 U\right)=\varepsilon^2 B r\left(\omega \rightarrow \pi^0 \gamma\right) \times \nonumber \\
& \hspace{1.5cm}\frac{\left[\left(m_\omega^2-\left(M_U+m_\pi\right)\right)\left(m_\omega^2-\left(M_U-m_\pi\right)\right)\right]^{3 / 2}}{\left(m_\omega^2-m_\pi^2\right)^3}, \\
& B r\left(K^{+} \rightarrow \pi^{+} U\right)=\frac{\alpha \varepsilon^2}{2^{10} \pi^2 \Gamma_T(K)} \frac{M_U}{m_K} W\left(M_U,m_K\right) \times \nonumber \\
& 
\hspace{5cm} \lambda^{1 / 2}\left(M_U, m_K, m_\pi\right),\\
& B r(V \rightarrow U)=\frac{\alpha \varepsilon^2 M_U}{3 \Gamma_T(V)}, \quad V=\rho, \phi, \omega,
\end{align}
where   $m_i$ corresponds to the mass of the particle $i$, 
$\lambda$ is the triangle function 
($\lambda(x,y,z)=(x-y-z)^2-4yz$) from the expression of particle 3-momentum, $A(m_\Delta)$ is the $\Delta$ spectral function, $\alpha$ is the fine structure constant, $\Gamma_T$ is the total width and $W=10^{-12}(3+6M_U^2/m_K^2)$. 
The branching ratio for the decay of $U$-bosons to $e^+e^-$ entering Eq. (\ref{NUee}) is 
adopted from Ref. ~\citep{Batell:2009yf} and used also in Ref. ~\citep{HADES:2013nab}:
\begin{eqnarray}
   Br^{U\to ee} &&
    = \frac{1}{1 + \sqrt{1 - \frac{4m_\mu^2}{M_U^2}} \left( 1 + \frac{2m_\mu^2}{M_U}\right) \left(1 + R(M_U)\right)},\label{Bree}
\end{eqnarray}
where $m_\mu$ is the muon mass. The total decay width of a $U$-boson is the sum of the partial decay widths 
to hadrons, $e^+e^-$, and $\mu^+\mu^-$ pairs,
$\Gamma_{tot}^U= \Gamma_{U\to hadr} + \Gamma_{U\to e^+e^-} + \Gamma_{U\to\mu^+\mu^-}$.
The expression (\ref{Bree}) has been evaluated using that $\Gamma_{U\to\mu^+\mu^-} = \Gamma_{U\to e^+e^-}$ 
due to lepton universality for $M_U\gg 2m_\mu$. The hadronic decay width of $U$-bosons 
is chosen such that $\Gamma_{U\to hadr} = R(\sqrt{s}=M_U)\Gamma_{U\to\mu^+\mu^-}$, where 
the factor $R(\sqrt{s}) = \sigma_{e^+e^-\rightarrow hadrons}$/$\sigma_{e^+e^-\rightarrow \mu^+\mu^-}$.
Eq. (\ref{Bree}) is only valid for $M_U<0.6$ GeV, and an extension of the branching ratio $Br^{U\to ee}$ up to $M_U<2$ GeV is taken from ~\citep{Liu:2014cma}.

\section{Results for the dilepton spectra from U-boson decays and constraints on $\epsilon^2 (M_U)$}
Given that both the kinetic mixing parameter, $\epsilon^2$, and the mass of the $U$-boson are not yet determined, we utilize the following approach to impose constraints on $\epsilon^2(M_U)$. For each dilepton mass bin $dM$, chosen as 10 MeV in our simulations, we compute the integrated dilepton yield from $U$-bosons with masses within the range $[M_U, M_U + dM]$ using Eq. (\ref{NUee}) and normalize it by the bin width $dM$. The resulting dilepton yield per mass bin $dM$ is denoted as $dN^{sumU}/dM$, representing the sum of all kinematically accessible dilepton contributions from $U$-bosons produced via the dark photon production channel $h \to \gamma X$.

Assuming that $\epsilon^2$ remains constant across $dM$, we express the dilepton yield as 
$dN^{sumU}/dM = \epsilon^2 \cdot dN^{sumU}_{\epsilon=1}/dM$, where $dN^{sumU}_{\epsilon=1}/dM$ represents the dilepton yield computed with $\epsilon = 1$.
The total yield from all possible dilepton sources, including both SM channels and $U$-boson decays, can then be written as:
\begin{equation}
\footnotesize
  \frac{dN}{dM}^{total} = \frac{dN}{dM}^{sum SM} + \frac{dN}{dM}^{sum U} =
    \frac{dN}{dM}^{sum SM} + \epsilon^2\frac{dN_{\epsilon=1}^{sum U}}{dM}.
\label{dNdMepsil}
\end{equation}
We can set constraints on $\epsilon^2(M_U)$ by requiring that the total dilepton yield, $dN^{total}/dM$, does not exceed the Standard Model contribution by more than a specified fraction $C_U$ for each mass bin $dM$. The parameter $C_U$ defines the maximum permissible increase in dilepton yield from dark photons compared to the Standard Model yield (for instance, setting $C_U = 0.1$ means that dark photons can contribute up to an additional 10\% to the yield observed from Standard Model processes). This condition can be formulated as
$ 
  dN^{total}/dM = (1+C_U) dN^{sum SM}/dM.
$ 
By combining the previous equation with Eq. (\ref{dNdMepsil}), it is possible to express the kinetic mixing parameter $\epsilon^2$ for a given $M_U$ as:
\begin{eqnarray}
    \epsilon^2 (M_U) = C_U \cdot  \left. { \left(\frac{dN}{dM}^{sumSM} \right)} \right/
    {\left(\frac{dN_{\epsilon=1}^{sum U}}{dM} \right)}.
\label{epsM}
\end{eqnarray}

Eq. (\ref{epsM}) offers a method to calculate $\epsilon^2$ for each mass interval $[M_U, M_U + dM]$, which represents the weighted dilepton yield from dark photons in comparison to the contributions from Standard Model processes. This approach also accounts for the experimental acceptance of $e^+e^-$ pairs produced by $U$-boson decays, treating them similarly to the contributions from Standard Model channels, ensuring compatibility with experimental data. Through a comparison of theoretical predictions with experimental measurements, the possible values for $C_U$, which regulates the extra yield from dark photons, can be determined. Given that dark photons have not been observed in dilepton experiments, any additional yield from these particles must remain within the uncertainties of the experimental data, under the assumption that Standard Model predictions are in agreement with the measurements.

%
%
\begin{figure*}[!]
\centerline{
\includegraphics[width=8.1 cm]{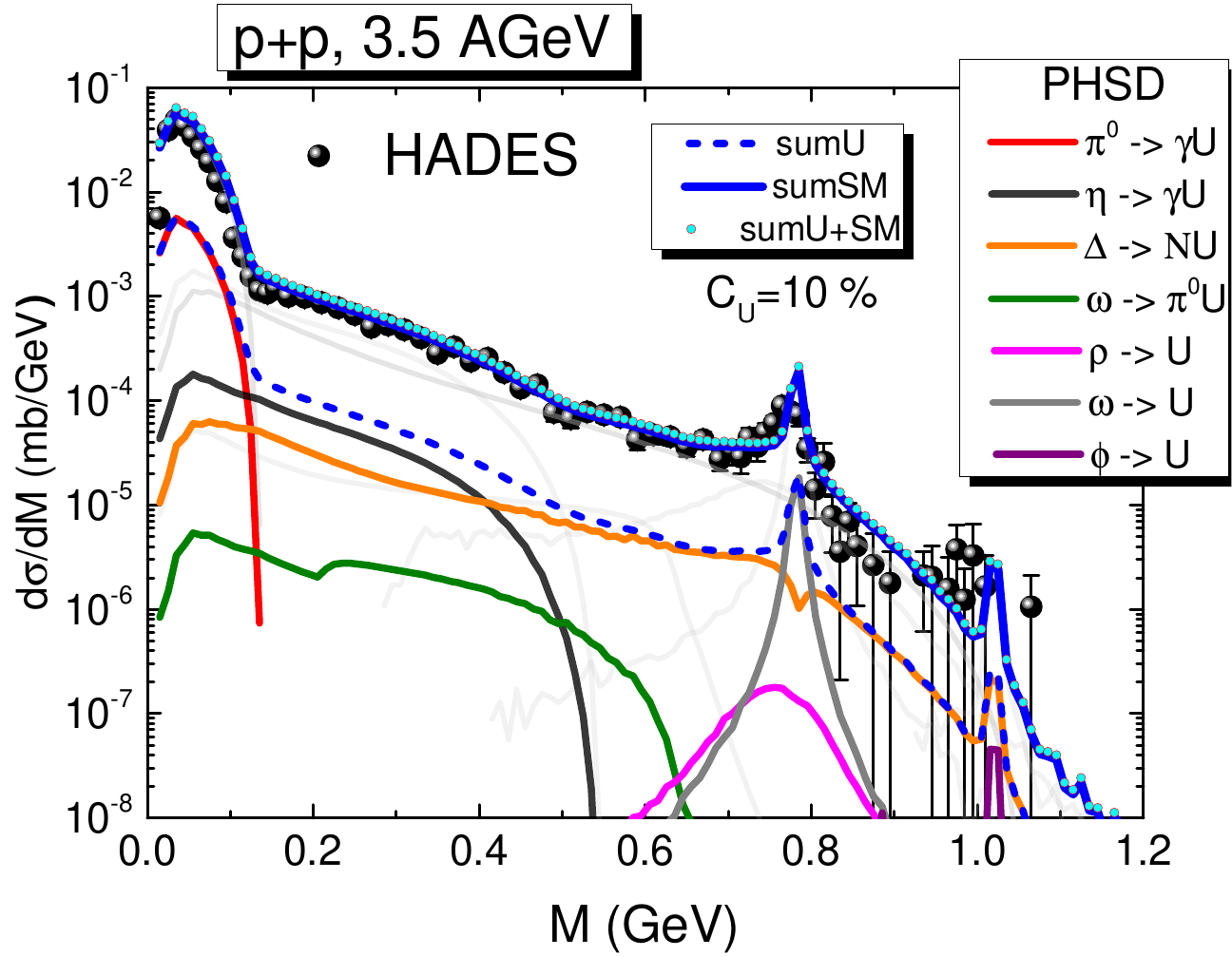}\hspace*{5mm}
\includegraphics[width=8.1 cm]{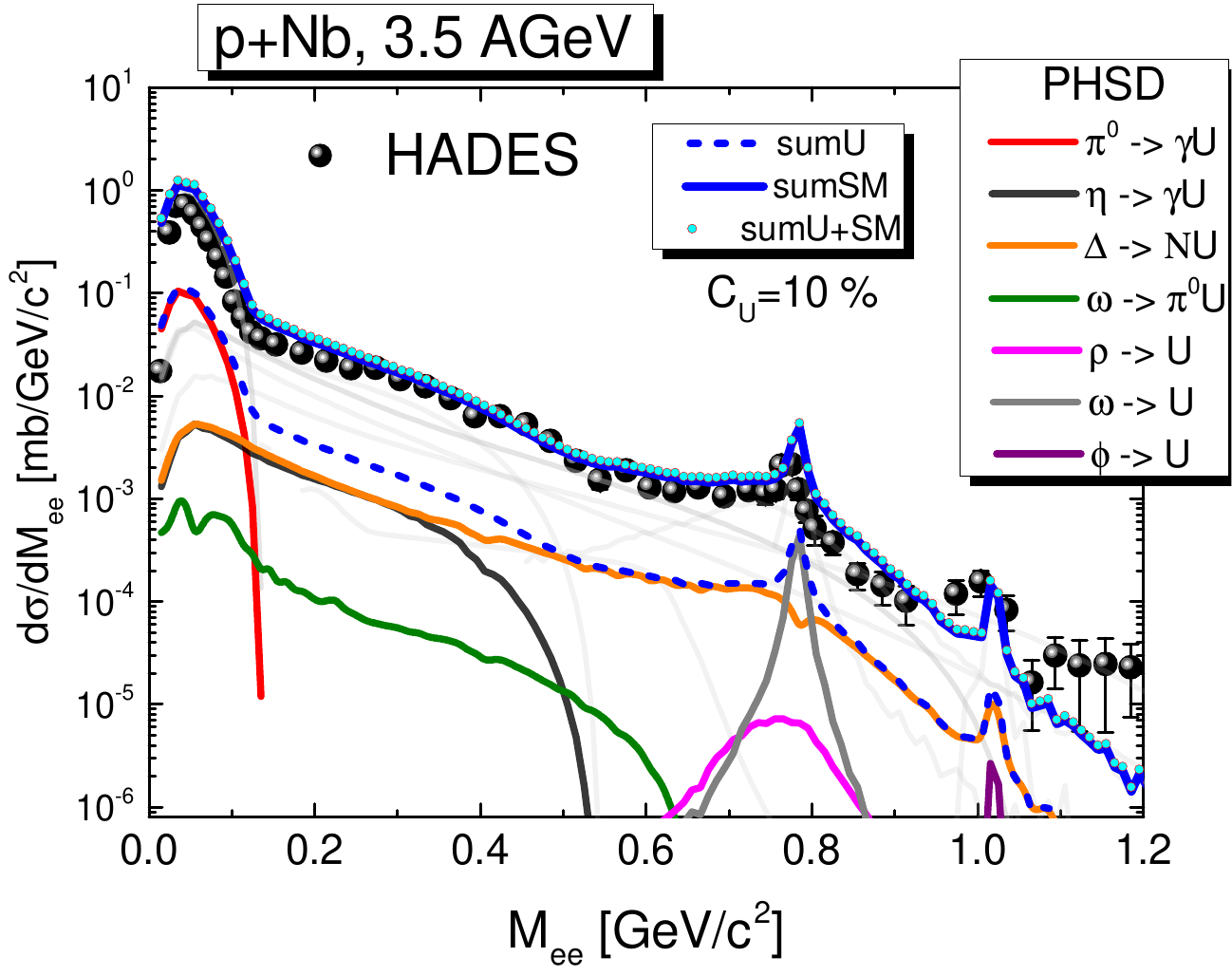}}
\centerline{
\includegraphics[width=8.1 cm]{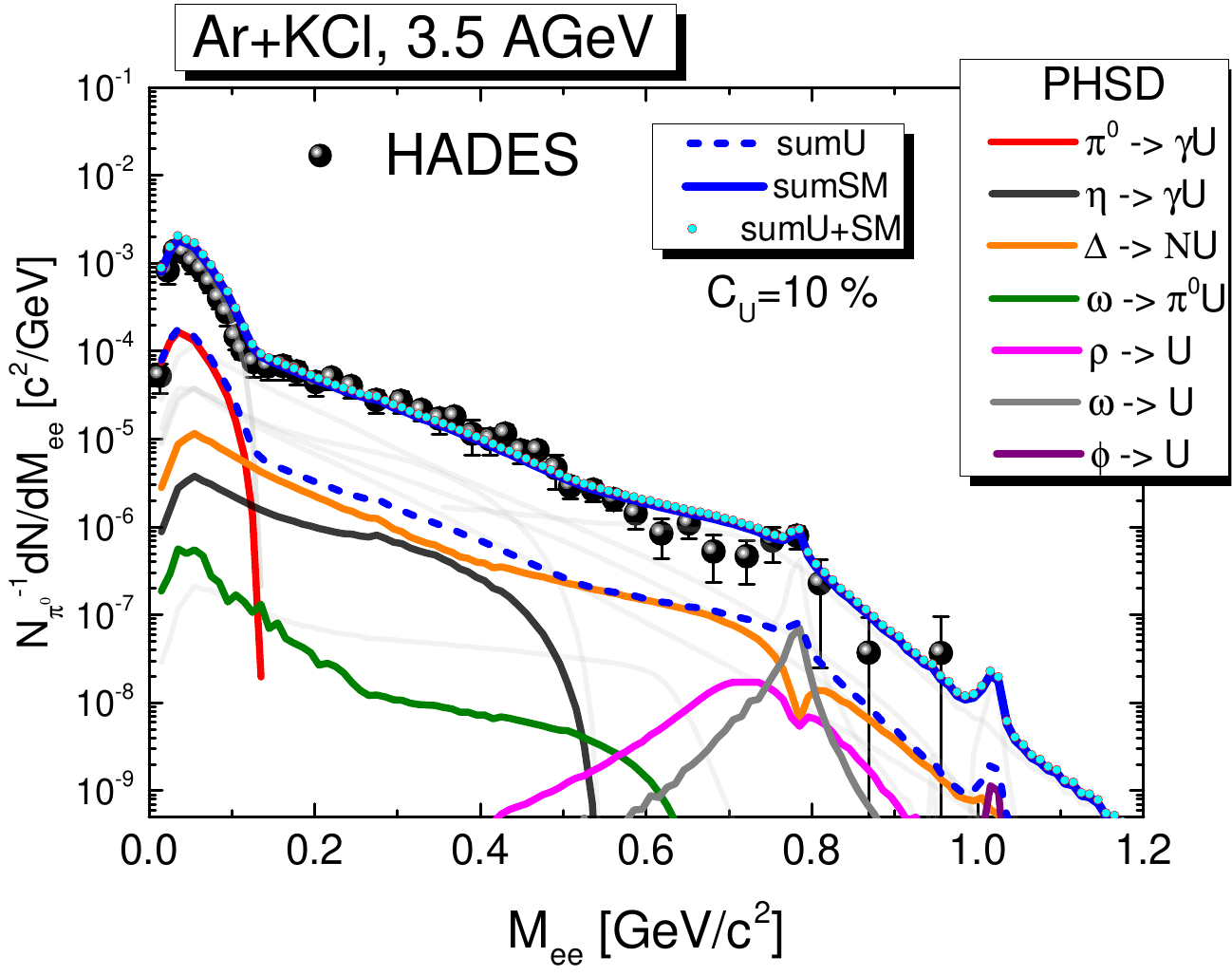}\hspace*{5mm}
\includegraphics[width=8.1 cm]{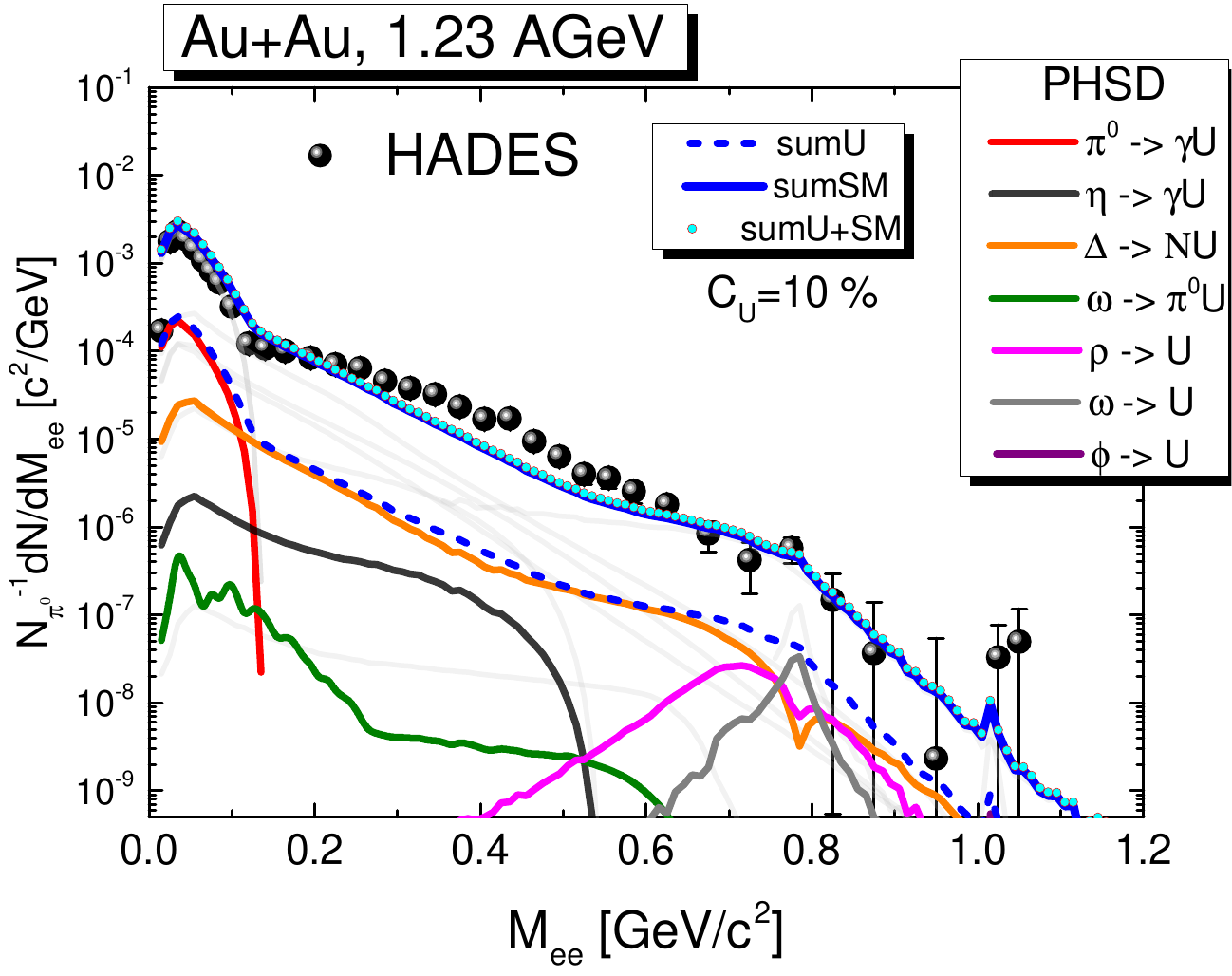}}
\caption{The PHSD simulations predict the differential cross section $d\sigma/dM$ for $e^+e^-$ production in $p+p$ (top-left panel) and $p+Nb$ collisions (top-right panel) at a beam energy of 3.5 GeV. Additionally, the calculations provide the mass differential dilepton spectra $dN/dM$, normalized to the $\pi^0$ multiplicity, for $Ar+KCl$ collisions at 1.76 A GeV (bottom-left panel) and for $Au+Au$ collisions at 1.23 A GeV (bottom-right panel). These results are compared with experimental data from the HADES Collaboration.
The light gray lines denote the SM channels included in the PHSD calculations.
The HADES measurements are represented by solid dots: $p+p$ data are taken from Ref. \cite{HADES:2011ab}, $p+Nb$ results from Refs. \cite{Weber:2011zze,Agakishiev:2012vj}, $Ar+KCl$ data from Ref. \cite{Agakishiev:2011vf}, and $Au+Au$ data from Ref. \cite{Adamczewski-Musch:2019byl}. The various SM contributions to dilepton production in PHSD are illustrated by individual colored lines, with the specific channels indicated in the legend.
Dileptons arising from $U \rightarrow e^+e^-$ processes, allowing for a 10\% surplus over the total SM yield, are included.
Dark photon contributions are categorized by their sources: Dalitz decays of $\pi^0$ (red), $\eta$ (black), $\Delta$ resonances (orange), $\omega$ (olive) and direct vector meson decays $\rho, \omega, \phi$ (magenta, dark yellow, purple) respectively. The combined yield from these decays is shown as a dashed blue line, while the total dilepton yield—including both SM and $U$-boson contributions—is represented by cyan dots.
We used the HADES acceptance criteria, incorporating its mass and momentum resolution. }
\label{M_Hades}
\end{figure*}
\begin{figure*}[!]
    \centering
     \includegraphics[width=0.44\linewidth]{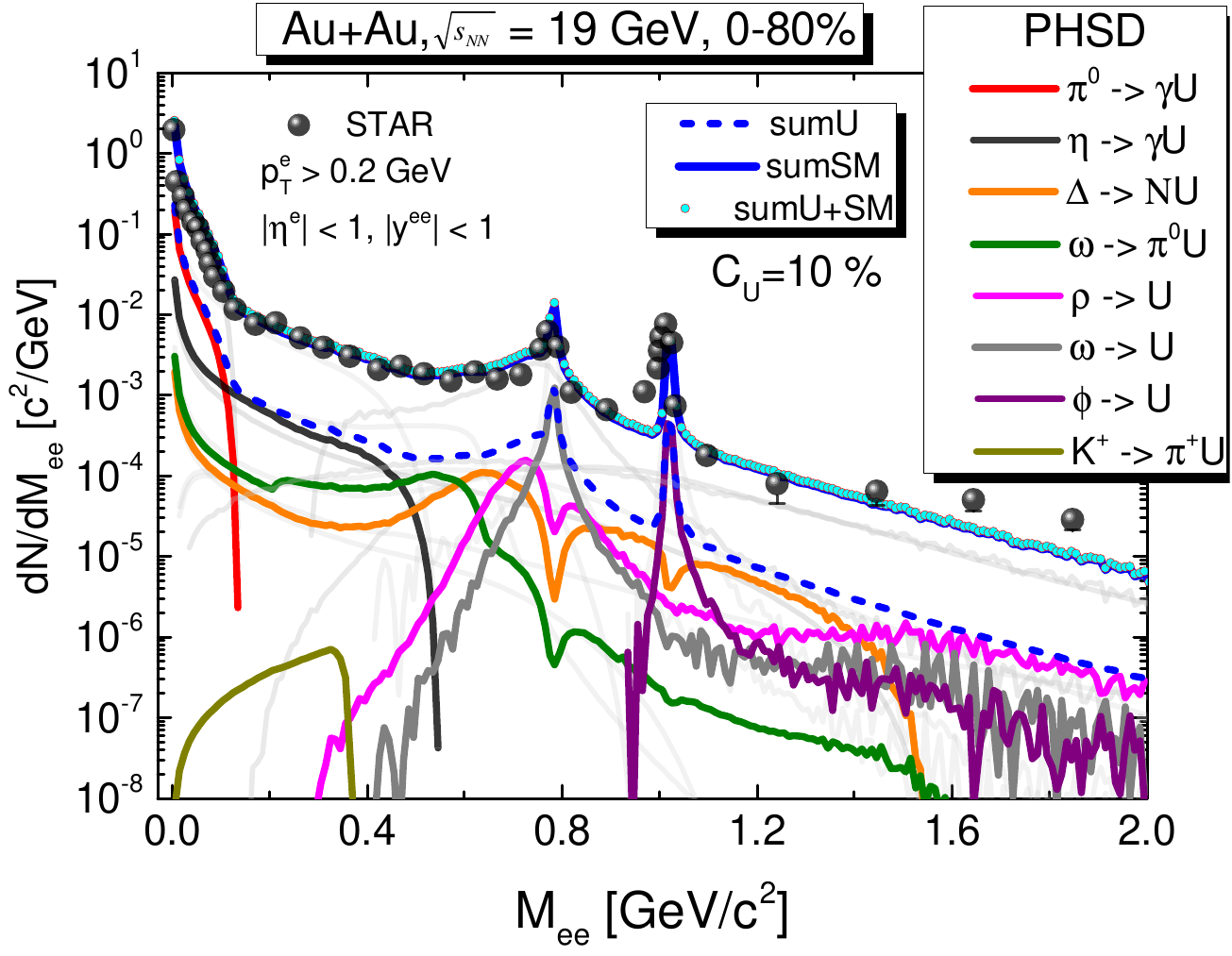}
     \includegraphics[width=0.42\linewidth]{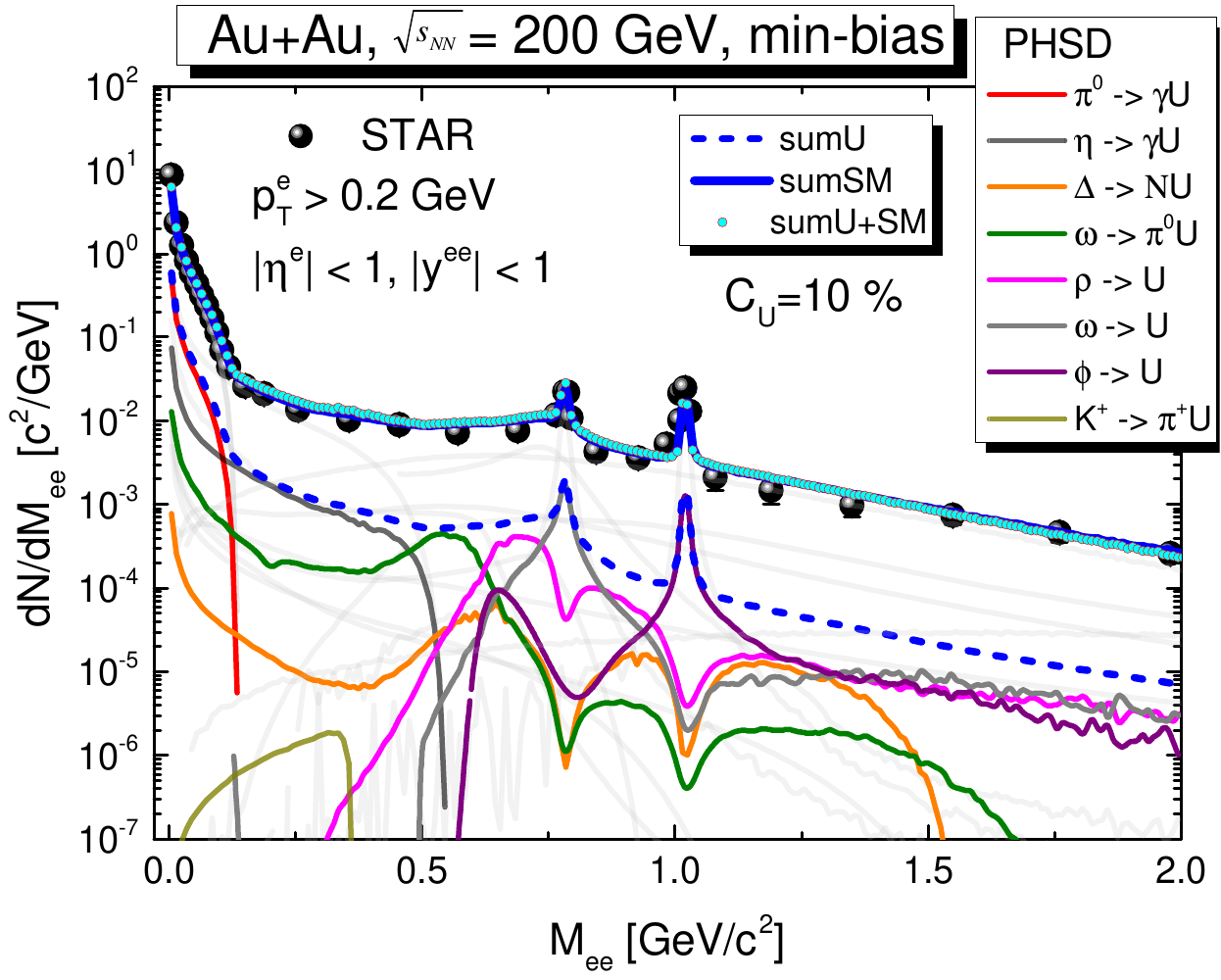}
    \caption{
The invariant mass spectra of dileptons produced in Au+Au collisions at $\sqrt{s_{NN}} = 19.6$ GeV (left panel) and $200$ GeV (right panel) are calculated using PHSD and compared to STAR experimental data  ~\citep{Han:2024nzr} and   ~\citep{STAR:2015tnn} respectively. The total dilepton yields predicted by the PHSD model are represented by blue lines, while the individual contributions from different production channels are detailed in the legends.
The light gray lines denote the SM channels included in the PHSD calculations. 
Dark photon contributions are categorized by their sources: Dalitz decays of $\pi^0$ (red), $\eta$ (black), $\Delta$ resonances (orange), $\omega$ (olive) and direct vector meson decays $\rho, \omega, \phi$ (magenta, dark yellow, purple) respectively.
In addition, the solid lines reflect the inclusion of $U \rightarrow e^+e^-$ decays, allowing for a 10\% surplus over the total SM yield.
The combined yield from these decays is shown as a dashed blue line, while the total dilepton yield—including both SM and $U$-boson contributions—is represented by cyan dots.
The STAR experimental data points are depicted as solid black dots.
To facilitate a direct comparison, the theoretical calculations are adjusted to align with the STAR acceptance criteria, incorporating its mass and momentum resolution.    }
    \label{RHIC_AuAu}
\end{figure*}
\begin{figure*}
\includegraphics[width=0.44\textwidth]{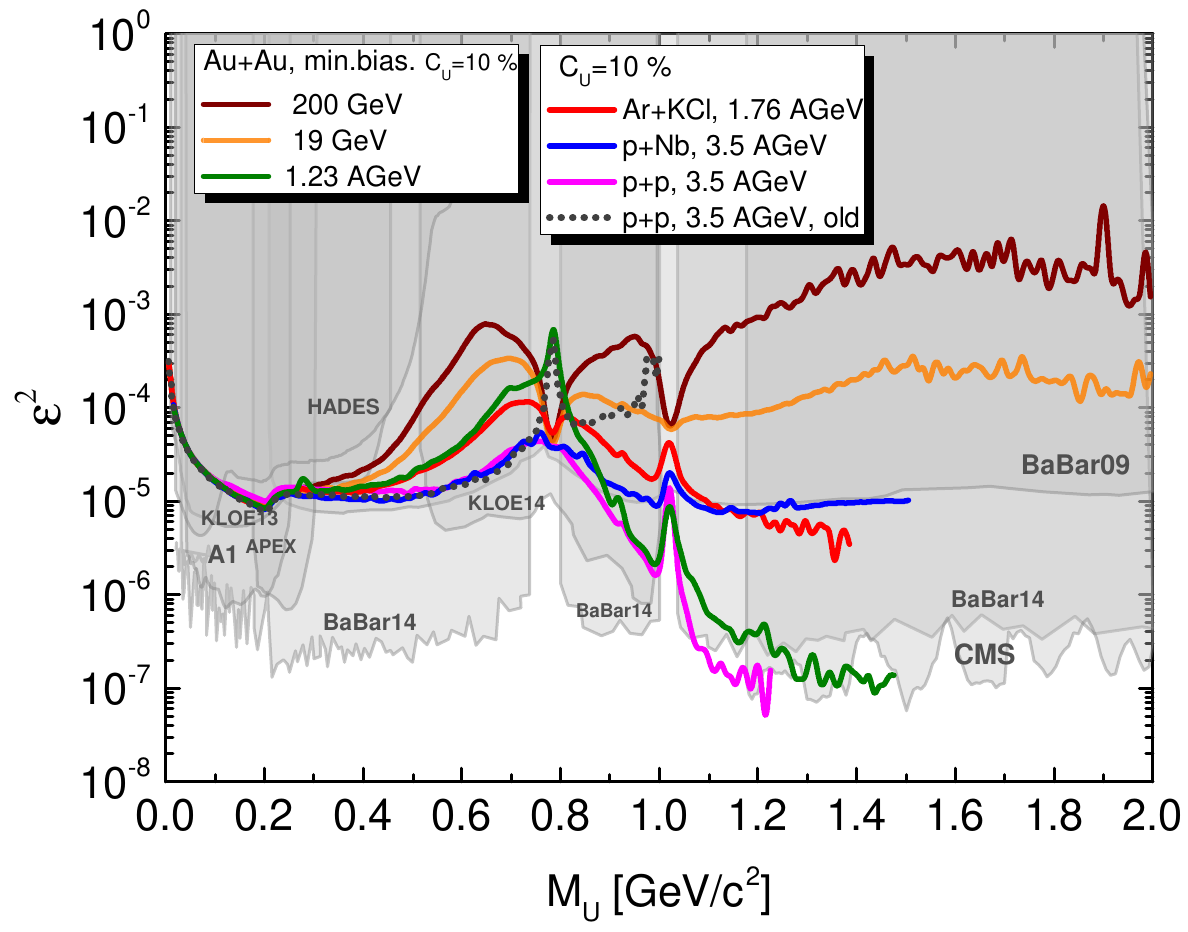}
\includegraphics[width=0.44\textwidth]{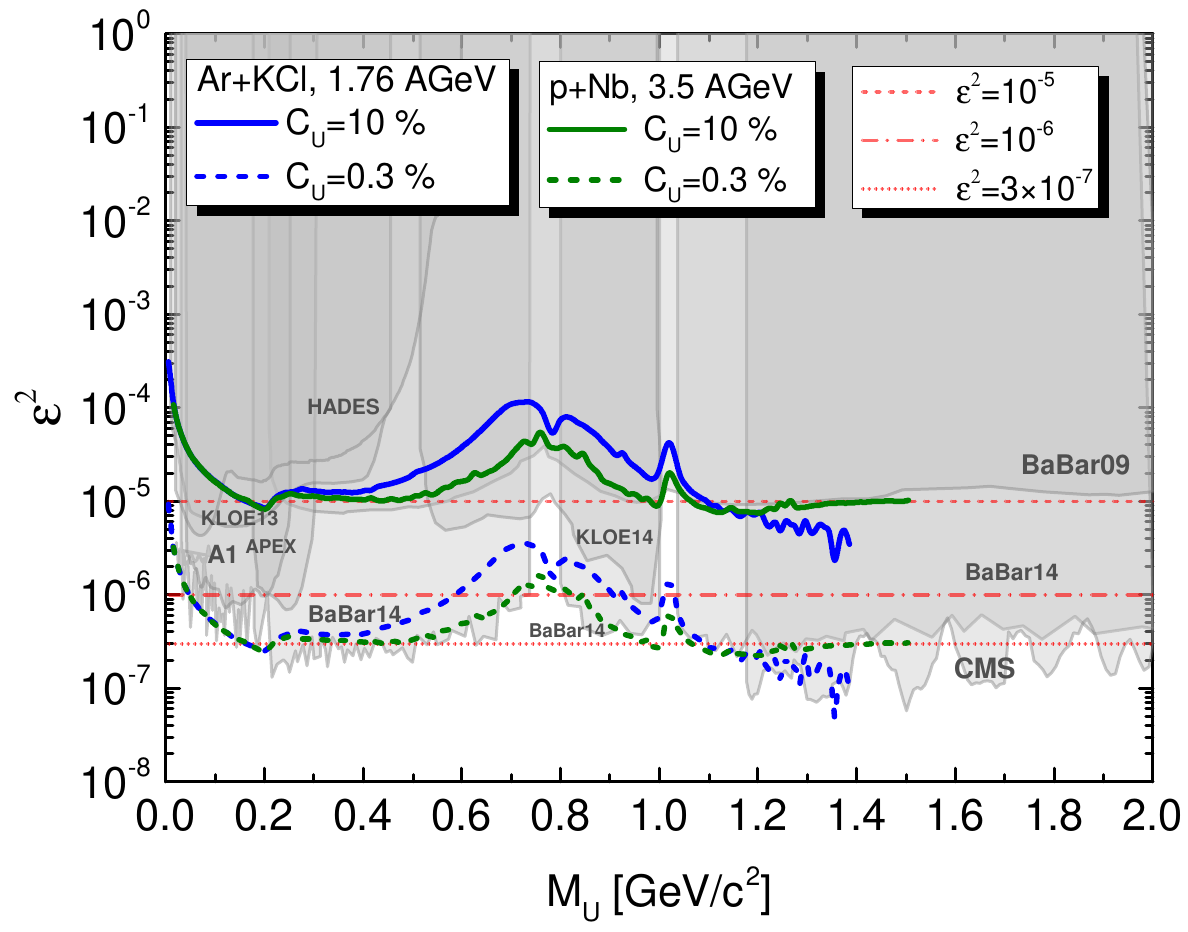}
\caption{
\textbf{Left panel}: The kinetic mixing parameter $\epsilon^2$ extracted from the PHSD dilepton spectra for $p+p$ at 3.5 AGeV (blue line), $Au+Au$ at 1.23 AGeV (olive line), $p+Nb$ at 3.5 AGeV (blue line),  $Ar+KCl$ at 1.76 AGeV (red line), $Au+Au$ at 27 GeV (red line) and 200 GeV (brown line). We also display experimental data for comparison with the existing exclusion limits at 90\% CL from HADES \cite{HADES:2013nab}, APEX \cite{APEX:2011dww}, A1 at MAMI \cite{Merkel:2014avp}, KLOE \cite{KLOE-2:2014qxg}, BaBar \cite{BaBar:2009lbr,BaBar:2014zli} and CMS \cite{CMS:2023hwl}.
The preliminary PHSD results are shown with 10\% allowed surplus of the $U$-boson contributions over the total SM. We include our previous result in ~\citep{Schmidt:2021hhs,Bratkovskaya:2022cch} for $p + p$ collisions at 3.5 GeV (dotted black line) for comparison with the current results.
\textbf{Right panel}: The kinetic mixing parameter $\epsilon^2$ extracted from the PHSD dilepton spectra for $Ar+KCl$ at 1.76 AGeV (blue) and $p+Nb$ at 3.5 AGeV (olive) calculated for different $\epsilon^2$ scenarios.
The PHSD results are shown with 0.3\% (dashed line) and 10\% (solid line) allowed surplus of the $U$-boson contributions over 
the total SM yield. The dashed, dotdashed, and dotted red lines shows, the constant $\epsilon^2 =10^{-5}$, $\epsilon^2 =10^{-6}$ and $\epsilon^2 =3 \times 10^{-7}$ respectively. 
}
\label{epsil2}
\end{figure*}
%

In Fig. \ref{M_Hades}, we display the PHSD results for the differential cross section $d\sigma/dM$ corresponding to $e^+e^-$ production in $p+p$ (top left) and $p+Nb$ reactions (top right) at a beam energy of 3.5 GeV. Additionally, we present the dilepton mass spectra $dN/dM$, normalized to the $\pi^0$ multiplicity, for $Ar+KCl$ collisions at 1.76\,A GeV (bottom left) and for $Au+Au$ collisions at 1.23\,A GeV (bottom right), alongside experimental data obtained by the HADES Collaboration. 
The contributions from different Standard Model (SM) channels of dilepton production, as calculated within the PHSD framework, are depicted as a gray line background while the blue solid lines indicate the sum of all SM contributions, labeled as "Sum SM" in the legend.

As illustrated in Fig.~\ref{M_Hades}, the PHSD calculations for SM channels exhibit strong agreement with HADES experimental data across all four systems under investigation. This alignment is consistent with earlier results reported using the PHSD approach in Ref.~\cite{Bratkovskaya:2013vx}.

The contributions from $U$-boson decays to dileptons are represented in Fig.~\ref{M_Hades} using various symbols. Specifically, the dileptons resulting from $U \to e^+e^-$ decays following Dalitz decays of $\pi^0$ (red), $\eta$ (black), $\Delta$ resonances (orange), $\omega$ (olive) and direct vector meson decays $\rho, \omega, \phi$ (magenta, dark yellow, purple) respectively. The total contribution from all kinematically allowed dark matter (DM) channels within a specific mass bin $M = M_U$ is illustrated using a dashed blue line, labeled as "Sum U" in the legend. Additionally, the combined yield from all Standard Model (SM) channels and $U$-boson decays is represented by cyan dots, marked as "Sum SM+U" in the legend.

The dilepton yields from $U$-boson decays have been computed by applying a surplus factor $C_U = 0.1$, corresponding to an allowance of a 10\% surplus over the theoretical predictions from the sum of all SM channels. This factor quantifies the additional contribution permitted from $U$-bosons relative to the Standard Model sources.

The production of dark photons at masses greater than 1 GeV is also of interest. Therefore, we extended our calculations to the Au+Au system at RHIC energies, where dark photon production increases with rising energy.

In Fig. \ref{RHIC_AuAu}, we display the dilepton mass spectra $dN/dM$ for $Au+Au$ collisions at 19 GeV (left panel) and 200 GeV (right panel) in comparison with the experimental data obtained by the STAR Collaboration. 
The light gray lines represent the Standard Model (SM) channels included in the PHSD calculations. The dark photon contributions are the same as in the before Fig. \ref{M_Hades}, including the kaon decay $K^+ \to \pi^+ U$ (dark yellow), which for RHIC energies takes a non-neglectable contribution.   
Furthermore, the solid lines account for the inclusion of $U \to e^+e^-$ decays, which allow for a 10\% surplus over the total SM yield.
The combined contribution from these decays is shown as a dashed blue line, while the total dilepton yield—considering both SM and $U$-boson contributions—is represented by cyan dots.
The experimental data from STAR are shown as solid black dots. For comparison, the theoretical calculations are adjusted to match the STAR acceptance, incorporating the corresponding mass and momentum resolution.

As shown in Fig.~\ref{RHIC_AuAu}, the PHSD calculations for both the Standard Model and U-boson channels are in excellent agreement with the STAR measurements. Notably, for high invariant mass $M > 1.2$ GeV, the dominant contributions to dark photon production come from vector mesons,($\rho$, $\phi$, $\omega$), and $\Delta$ resonances.

As observed in Figs.~\ref{M_Hades} and \ref{RHIC_AuAu}, the overall contributions from dark matter (DM) sources closely follow the shape of the corresponding Standard Model (SM) distributions. This behavior arises from the constraints on $\epsilon^2(M_U)$ imposed by Eq.~(\ref{epsM}), which effectively scales the DM yield by the same factor $C_U$ across all bins of $M_U$. 

The significance of vector mesons in dilepton production becomes increasingly pronounced at higher beam energies, particularly for masses exceeding $M > 0.6$ GeV, as highlighted in Refs.\cite{Batell:2009yf, Batell:2009di}. $U$-bosons generated via vector meson conversion processes play a crucial role in accurately determining the kinetic mixing parameter.

In  Fig. \ref{epsil2}, we show the (preliminary)  results for the kinetic mixing parameter $\epsilon^2$ versus $M_U$ extracted from the PHSD dilepton spectra for $p+p$ at 3.5 AGeV (blue line), $Au+Au$ at 1.23 AGeV (olive line), $p+Nb$ at 3.5 AGeV (blue line),  $Ar+KCl$ at 1.76 AGeV (red line), $Au+Au$ at 27 GeV (red line) and 200 GeV (brown line) in comparison with the world data compilation (gray background) ~\citep{Battaglieri:2017aum}.

The PHSD results are shown with a 10\% allowed surplus of the $U$-boson contributions over the total SM yield in the left panel of Fig.  \ref{epsil2}  for the systems mentioned previously. 
We present our calculations for larger $M_U$ values to illustrate that our theoretical approach can yield valuable constraints on $\epsilon^2$ across a range of $M_U$.
It is important to note that the shape of the theoretically derived $\epsilon^2(M_U)$ curve remains unaffected by the experimental detector acceptance, as this factor influences both SM and DM contributions equally at a given mass $M = M_U$. For smaller $M_U$ values (i.e., $M_U < m_{\pi^0}$), the extracted $\epsilon^2$ shows minimal dependence on the collision system size and energy, primarily because the Dalitz decay of $\pi^0$ is the dominant decay channel. However, as $M_U$ increases, additional decay channels become accessible, and the resulting constraints on $\epsilon^2$ are influenced by the proportion of $U$-boson production channels relative to all other dilepton production channels. We found that the extracted upper limit of $\epsilon^2(M_U)$ for $p+p$ at 3.5 AGeV is consistent with the results of the BaBar 2009 experiment between 0.2 < $M_U $ < 1 GeV with $C_U=10\%$. Furthermore, for $p+Nb$ at 3.5 AGeV, the kinetic mixing parameter is even better in agreement with the BaBar 2009 experiment for $0.2 < M_U < 1.5$ GeV. 

For invariant mass regions exceeding $M > 1$ GeV, the results for $Au+Au$ collisions at 19 and 200 GeV deviate from the expected values due to the influence of the quark-gluon plasma (QGP) and charm production in the left panel of Fig.  \ref{epsil2}.
Since our model does not account for the production of dark photons at high invariant masses, the contributions from the QGP and charm production dominate in this region. It is worth highlighting that, as suggested in Refs.~\cite{Fabbrichesi:2020wbt,Alexander:2016aln}, $U$-bosons can also be produced via Drell-Yan processes ($q\bar{q} \to U$) and Bremsstrahlung mechanisms ($e^-Z \to e^-Z U$). These production channels are not included in the current study, but represent an important avenue for future research.
In contrast, for low-energy systems (SIS18), where QGP and charm contributions are negligible, the model yields significantly better agreement, as illustrated in the graph for $p+p$, $Ar+KCL$ and $p+Nb$.


In the right panel of Fig.~\ref{epsil2}, the kinetic mixing parameter $\epsilon^2$ extracted from the PHSD dilepton spectra is displayed. The results are presented for various surplus values of the $U$-boson contributions relative to the total Standard Model (SM) yield, as denoted by the color coding in the legend.
Additionally, the dashed, dot-dashed, and dotted red lines represent constant values of $\epsilon^2 = 10^{-5}$, $\epsilon^2 = 10^{-6}$, and $\epsilon^2 = 3 \times 10^{-7}$, respectively. The latter value approximately corresponds to the current upper limit established by the compilation of global experimental data \cite{Agrawal:2021dbo}.


However, with the increasing precision of experimental data over the years, the upper limit of $\epsilon^2(M_U)$ has evolved, as evidenced by the BaBar 2014 results. By adjusting the dark photon surplus, we can refine our estimation to align with the experimental data, concluding that a dark photon surplus of $C_U=0.3$\% is necessary for consistency with the BaBar 2014 experimental data for $p+Nb$ at 3.5 AGeV and $Ar+KCl$ at 1.76 AGeV, which correspond approximately in the current upper limit of a  constant $\epsilon^2 = 3 \times 10^{-7}$ (dotted red line) from 0.2 < $M_U $ < 1.2 GeV.

%
\section{Summary}
This work presents a microscopic transport study employing the PHSD framework to evaluate the dilepton yields originating from hypothetical dark photons (or $U$-bosons) decaying into $e^+e^-$ pairs in $p+p$ and $A+A$ collisions at relativistic energies. It builds upon the previous analysis in Ref.~\citep{Schmidt:2021hhs,Bratkovskaya:2022cch}, which focused on dark photon production through $\pi^0 \to \gamma + U$, $\eta \to \gamma + U$, and $\Delta \to N + U$ channels, by incorporating additional processes. These include the direct decay of vector mesons ($V \to U$, where $V = \rho, \phi, \omega$), the $\omega$ Dalitz decay ($\omega \to \pi^0 + U$), and the kaon decay ($K^+ \to \pi^+ + U$).

To establish theoretical limits on the kinetic mixing parameter $\epsilon^2(M_U)$, we adopt the methodology outlined in Ref.~\citep{Schmidt:2021hhs,Bratkovskaya:2022cch}. Given that dark photons have not been detected in dilepton measurements to date, the approach enforces a constraint whereby their contribution must not exceed a predefined surplus level ($C_U$), which would render them detectable in the experimental data.

The extracted upper limit of $\epsilon^2(M_U)$ for $p+p$ collisions at 3.5 AGeV aligns well with the experimental results from the BaBar09 experiment in the mass range $0.2 < M_U < 1$ GeV with a surplus factor of $C_U = 10\%$. Additionally, for $p+Nb$ collisions at 3.5 AGeV, the kinetic mixing parameter demonstrates even closer agreement with the BaBar09 data, extending over the range $0.2 < M_U < 1.5$ GeV.
Similarly, the results are consistent with the BaBar14 experiment for the range $0.2 < M_U < 1.5$ GeV when $C_U = 0.3\%$, suggesting a possible dark photons contribution of 0.3\% of the SM yield. These outcomes also agree with the global compilation of experimental constraints on $\epsilon^2(M_U)$.
%
\vspace*{1mm}

\section*{Acknowledgements}
A.R.J. expresses gratitude for the financial support from the Stiftung Giersch. We also acknowledge the support by the Deutsche Forschungsgemeinschaft (DFG) through the grant CRC-TR 211 "Strong-interaction matter under extreme conditions" (Project number 315477589 - TRR 211) and the CNRS Helmholtz Dark Matter Lab (DMLab). The computational resources utilized for this work were provided by the Center for Scientific Computing (CSC) at Goethe University Frankfurt.
%
\bibliography{references}%
\end{document}